\documentstyle[12pt,aaspp4,epsf]{article}

\lefthead{Merritt \& Oh}
\righthead{Dynamics of M87}

\begin{document}

\title{The Stellar Dynamics of M87}

\author{David Merritt}
\affil{Department of Physics and Astronomy, Rutgers University,
    New Brunswick, NJ 08855}
\author{Siang Peng Oh}
\affil{Princeton University Observatory, Princeton, NJ 08544}
\bigskip
\centerline{Rutgers Astrophysics Preprint Series No. 195}
\bigskip

\begin{abstract}
We extract the shape of the stellar velocity ellipsoid as a function 
of radius in M87 from van der Marel's (1994) velocity dispersion data.
We include the gravitational force of a central black hole 
with the mass quoted by \cite{har94}.
The kinematical data are corrected for the effects of seeing and 
instrumental blurring using a nonparametric algorithm.
We find that the stellar motions in M87 are slightly radially 
anisotropic throughout the main body of the galaxy, 
with $\sigma_r\approx 1.2 \sigma_t$.
However $\sigma_t$ {\it exceeds} $\sigma_r$ within the inner 
$1''-2''$ by a statistically significant amount.
A number of models for the formation of nuclear black holes 
predict a tangential anisotropy in the stellar motions,
and our results provide evidence for such an effect in M87.
\end{abstract}

\section{Introduction}

Possibly no elliptical galaxy has been as thoroughly studied as M87 
(NGC 4486) in the Virgo cluster.
Early observations of the core of M87 revealed a velocity 
dispersion profile that rose toward the center, a 
feature that was initially attributed to the presence of a 
massive ($\sim 5\times 10^9 M_{\odot}$) black hole (Sargent et 
al. 1978).
The same data were soon interpreted in other ways; for instance, 
\cite{bin82} demonstrated that models without a central 
point mass were equally acceptable if one was willing to let the 
stellar velocity ellipsoid become strongly prolate at $\sim 100$ 
pc from the center.
Such extreme models were later shown to be dynamically unstable 
(Merritt 1987),
but the instability led only to the formation of a mildly 
elongated bar, which -- if viewed from a line-of-sight not far 
from its major axis -- yielded nearly the same dependence of velocity 
dispersion on radius as the anisotropic spherical model.
The most recent, stellar kinematical data (Dressler \& Richstone 
1990; Jarvis \& Peletier 1991; van der Marel 1994) reveal that 
the velocity dispersion fails to rise within the 
inner few arc seconds of M87.
These data require even less anisotropy than the earlier, 
black-hole-free models.
In fact, as emphasized by \cite{dre90}, a $\sim 
4\times 10^9M_{\odot}$ black hole can only be reconciled with the 
velocity dispersion data if one assumes that the stellar motions 
near the center of M87 are nearly {\it circular}, a configuration 
that they found implausible.
The stellar dynamical data thus seem to provide little if any 
evidence for a dark central mass.

The case for a massive black hole in M87 is nevertheless quite 
strong.
\cite{for94} and \cite{har94} discovered a rapidly-rotating, 
ionized gas disk within the inner arc second of M87, with a major axis 
roughly perpendicular to the M87 jet.
At a radius of $\pm0.22''=16$ pc along the major axis of the 
disk, the spectra show emission lines separated by $2V = 916 $ km 
s$^{-1}$.
If this gas is in approximately circular motion, and if the disk 
is inclined at $42^{\circ}$ as implied by the apparent axial 
ratio, then M87 contains a central dark mass of $M_{h}\approx 
2.4\times 10^9 M_{\odot}$ (Harms et al. 1994).
More recent HST observations at higher spatial resolution (Ford 
et al. 1996) appear to confirm this result.

A nuclear black hole with this mass would be expected to affect
the stellar motions within a radius $r_{h}=
GM_{h}/\sigma^2\approx 80\ {\rm pc} \approx 1''$ in M87, where 
$\sigma\approx 400$ km s$^{-1}$ is the stellar velocity dispersion.
(We assume throughout a distance to M87 of 16 Mpc; thus $1''$ 
corresponds to $77.6$ pc.)
However no increase in the stellar velocities is seen in the inner few
arc seconds; in fact the best ground-based 
data suggest a slight {\it decrease} in the velocity dispersion 
inside $1''-2''$ (e.g. Fig. 1 of Jarvis \& Peletier 1991).

These apparently contradictory data can be reconciled if
the stellar motions are appreciably circular 
within the radius of influence of the black hole.
Such a model will not keep the observed velocity dispersions low 
at arbitrarily small radii -- orbital velocities sufficiently close 
to the black hole must eventually rise according to Kepler's law -- 
but a local bias toward circular motions at $r\approx r_{h}$ can reduce 
the line-of-sight dispersions at $\sim 1''$ by an amount that is 
sufficient to reproduce the stellar velocity dispersion measurements
(Dressler \& Richstone 1990).
It is interesting that a number of models for the 
formation of massive black holes predict anisotropy
in the stellar motions at radii close to $GM_{h}/\sigma^2$.
For instance, the adiabatic growth of a black hole in a 
pre-existing galaxy core always produces a slight enhancement of 
circular motions at the expense of radial ones (Goodman \& Binney 
1984).
An even stronger effect may be predicted by models in which 
black holes form  
by the coalescence of smaller black holes (Ebisuzaki, Makino \&
Okumura 1991); here the anisotropy results from the ejection of 
stars on elongated orbits due to three-body scattering by the black
hole binary (Quinlan 1996).

In the present paper, we re-analyze the \cite{van94}
velocity dispersion data, assuming the presence of a nuclear 
black hole with the mass quoted by \cite{har94}.
We make use of the fact that -- in a spherical system with a 
known gravitational potential -- the shape of the 
stellar velocity ellipsoid at every radius follows uniquely from 
the observed velocity dispersion profile (Binney \& Mamon 1982).
In the case of M87, the stellar $M/L$ (e. g. Richstone \&
Tremaine 1985) implies that the black hole dominates the 
gravitational force out to a radius of $\sim 250$ pc $\approx 3''$.
The stars also contribute to the gravitational force, of course,
but their contribution can be reasonably approximated by assuming 
a constant mass-to-light ratio; $M/L$ can then be derived from
the virial theorem independent of any assumptions about the 
velocity anisotropy.
Such a model is likely to be only approximately correct at large 
radii, where the dark matter surrounding M87 begins to affect
the gravitational potential.
But it is probably an excellent approximation at small radii 
where the influence of the black hole on the motions of the stars 
is significant.

The same conceptual model has been used in many previous studies of the 
stellar dynamics of M87.
Our treatment is new, however, in two ways.
First, ours is the first study to to make use of the
black hole mass as determined by Harms et al. (1994) and Ford et 
al. (1996).
Thus we are able to specify a definite value for $M_h$ (although
we consider also the effect of varying $M_h$ by its estimated 
uncertainty).
Second, our treatment of the data is more sophisticated than in 
previous studies.
Since the dependence of the velocity anisotropy on radius,
$\beta(r) = 1-\sigma_t^2(r)/\sigma_r^2(r)$, is not known a priori, 
the goal should be to generate an estimate of this function directly 
from the data.
In other words, one should approach the estimation of $\beta(r)$ as an 
inverse problem, not as a model-fitting problem.
\cite{bin82} attempted to infer $\beta(r)$ directly from the data
that were available at the time, but they chose to represent the
velocity dispersion profile using an ad-hoc set of template curves,
thus weakening the model-independence of their results.
Binney \& Mamon also did not take into account the possible effects of 
seeing and instrumental blurring on the velocity dispersion data.
Van der Marel (1994) took the opposite, ``model-fitting'' 
approach, assuming a parametrized functional form for $\beta(r)$ and 
then predicting how a given model would appear after projection into 
observable space.
However he did not consider any functional forms for $\beta(r)$ in 
which the stellar motions were circularly biased 
at small radii, as predicted by the black-hole formation models 
discussed above.

The algorithm presented here combines the best features of both 
these studies.
We ask simply: What smooth functions $\beta(r)$ are consistent with
the measured set of velocity dispersions?
This is an ill-posed problem, since many functional forms
for $\beta(r)$ are likely to be consistent with the data after 
projection through the galaxy and after blurring by the 
atmosphere and the detector.
We accordingly use a regularized algorithm to find the ``most 
likely'' expression for $\beta(r)$.
The uncertainty in our estimate is itself estimated via a standard 
bootstrap algorithm; the ill-conditioning of the inverse
problem manisfests itself most strongly as a widening of the
confidence bands at small and large radii, where the data do
not strongly constrain the solutions.
At no point do we make any assumptions about the functional forms 
of the observed profiles or of $\beta(r)$, except insofar as we 
require these functions to be continuous and smooth.
Thus we achieve the model-independence aspired to by \cite{bin82},
while fully accounting for observational degradation of the 
data as in \cite{van94}.

We find that a black hole with the mass determined by \cite{har94} 
and \cite{for96} implies a substantial change in the 
shapes of the stellar orbits near the center of M87.
In order to reproduce the low, central value of the velocity 
dispersion, the stellar orbits must become appreciably circular 
within the radius of influence of the black hole.
The statistical significance of this result depends somewhat on 
the exact value assigned to $M_h$; the larger $M_h$, the greater 
the tangential anisotropy implied by the velocity dispersion 
data.
Such a result -- although superficially implausible for a hot 
stellar system -- deserves to be taken seriously in the light of
black-hole formation models that predict a circular bias in the 
stellar orbits at $r\approx r_h$.

The data, none of which are new, are summarized in \S 2.
The nonparametric algorithm used to correct the velocity dispersion data 
for seeing and for instrumental blurring is presented in \S3.
The derivation of $\beta(r)$ is the subject of \S4, and in \S5 
we interpret our results in the light of recent models for the 
formation of nuclear black holes.

\bigskip

\section{Data}

Only two observational quantities are needed for this study: 
the surface brightness profile of the starlight in M87, $\Sigma(R)$; 
and the line-of-sight velocity dispersion profile of the stars,
$\sigma_p(R)$.
(We assume throughout that M87 is spherically symmetric.)
The published sources from which our data were derived are as follows.

HST I-band photometry of M87 for $R<20''$ were presented by \cite{lau92}.
These data have an estimated spatial resolution of $\sim 0.1''$.
Between $20''$ and $163''$, ground-based, $R_c$ band photometry 
of M87 was published by \cite{pel90}.
We converted the latter data to I-band via $R_c-I=1.153$, 
the average color at these radii.
The surface brightnesses were then corrected for galactic 
extinction using $A_I=0.04$ (van der Marel 1994).

Our velocity dispersion data were taken from \cite{van94},
who used the 4.2m William Herschel Telescope in subarsecond seeing 
to obtain high S/N spectra of M87 within $\sim 25''$.
We followed van der Marel's practice of using only the G-band 
data presented in his Table C1.
Measurements of $\sigma_p$ at positive and negative radii were 
averaged and their quoted errors reduced using standard formulae 
for error propagation.
These data are presented in Fig. 1.

\bigskip

\section{Correction of the Data for Seeing and Instrumental Blurring}

The data consist of a set of measurements $d_i$ at discrete 
positions $R_i$ and their estimated errors $\epsilon_i$.
Our goal is to extract smooth representations of functions like 
$\beta(r)$ from these (noisy and incomplete) data.
The relation between the data and the unknown function (call it $u$) 
has the implicit form
\begin{equation}
d_i = {\cal L}_i u + e_i
\end{equation}
where ${\cal L}_i$ is a linear operator relating $u$ to 
measured quantities at $R=R_i$, and $e_i$ represents scatter of the data 
around their mean value at $R_i$, due in part to 
measurement errors.

The operator ${\cal L}$ contains contributions from two sorts 
of effects: data degradation due to instrumental limitations,
seeing, binning, etc; and projection of the intrinsic function $u$ 
into observable space -- for instance, projection onto the plane of the 
sky.
To the extent that both effects are understood, at least in a 
statistical sense, we can write an 
expression for ${\cal L}$ and apply regularized inversion 
algorithms to solve for the most likely $u$ (Merritt 1993a).
Confidence bands on the function estimates can then be computed via the 
bootstrap (Efron 1982).
The regularization is necessary since the inverse operation 
${\cal L}^{-1}$ is typically ill-conditioned (O'Sullivan 1986), 
so that direct inversion of the data would produce unacceptably 
noisy results.

Rather than derive a single expression for ${\cal L}$ that includes all
of the relevant effects, we chose first to correct the data for seeing and 
instrumental blurring, and then to operate directly on the resulting 
smooth profiles via the inverse operator ${\cal L}^{-1}$ that 
represents spatial deprojection alone.
This two-step procedure is perfectly justified from a statistical
point of view (e.g. Wahba 1990, p. 19) and has the advantage that
it allows us to separate the effects of instrumental blurring
from the effects of spatial deprojection.

The spatial resolution of the \cite{lau92} HST surface photometry, $\sim 
0.1''$, is much greater than that of the velocity dispersion 
observations in the region $R<20''$ where the two data sets 
overlap.
We can therefore ignore instrumental blurring in our estimate 
of the stellar luminosity density profile $\nu(r)$.
We derived a nonparametric estimate of 
$\Sigma(R)$ by fitting a smoothing spline (Wahba 1990) to the 
logarithms of the photometric data, $\log\Sigma_i$ vs. $\log R_i$.
The smoothing parameter was chosen by generalized 
cross-validation (Wahba 1990, p. 52).
This estimate was then substituted into Abel's equation,
\begin{equation}
\nu(r)=-{1\over\pi}\int_r^{\infty} {d\Sigma\over dR} 
{dR\over\sqrt{R^2-r^2}},
\label{Abel}
\end{equation}
to yield an estimate of $\nu(r)$.
The result was found to be almost indistinguishable from the profile
displayed by van der Marel (1994, Fig. 8), who used the same photometric 
data as us. 
We therefore do not reproduce our estimate of $\nu(r)$ here.
We note in passing that M87 -- although still often described as having a 
``core'' (e.g. Kormendy \& Richstone 1995) -- in fact has a 
luminosity density profile
that increases as a nearly-perfect power law at small 
radii, $\nu\sim r^{-1}$, $r<100$ pc.
This profile appears superficially core-like since
an $\sim r^{-1}$ density profile produces only a logarithmic 
divergence when seen in projection.

In the case of the velocity dispersion measurements, which were 
made from the ground, the observed values might be strongly affected 
near the galaxy center by atmospheric seeing and by the finite 
spatial resolution of the detector.
Here we generate, from the observed velocity dispersions $\sigma_p'(R_i)$,
an estimate of the true velocity dispersion profile $\sigma_p(R)$ 
that would have been observed using a perfect detector in the 
absence of atmospheric blurring.
This estimate will be used in the next 
section to generate estimates of the anisotropy profile.

The true velocity dispersion profile $\sigma_p(R)$ is related 
to the observed profile $\sigma_p'(R)$ via
\begin{eqnarray}
\Sigma'(R){\sigma_p'}^2(R) = {1\over4lw}
& & \int_{-\infty}^{\infty} dX 
\int_{-\infty}^{\infty} dY\ \Sigma(\sqrt{X^2+Y^2})\ {\sigma_p}^2
(\sqrt{X^2+Y^2}) \times \nonumber \\
& & \int_{R-l}^{R+l} dx\int_{-w}^{w}dy\ {\rm PSF}
\left[\sqrt{(X-x)^2+(Y-y)^2}\right]
\end{eqnarray}
(e.g. Qian et al. 1995).
Here PSF is the atmospheric point spread function; we adopt van der 
Marel's (1994) expression:
\begin{equation}
{\rm PSF}(R) = A_1e^{-R^2/2\sigma_1^2} + A_2e^{-R^2/2\sigma_2^2},
\end{equation}
with $\sigma_1=0.313''$, $\sigma_2=0.751''$, $A_1=0.929$, 
and $A_2=0.121$.  
The inner integration is over the rectangle whose $x$-dimension 
$2l$ is the CCD pixel width, and whose $y$-dimension $2w$ is the slit 
width.
These values were taken from Tables 1 and C1 of \cite{van94};
beyond $3''$ from the center, data from more than one pixel
were binned together, making $l$ a function of position.
The outer integration is over the 2-D image of the galaxy, assumed to 
be spherically symmetric.
Finally, $\Sigma'(R)$ is the surface brightness profile as it would
appear after convolution with the seeing disk and the detector.

Interchanging the order of integration, we can write
\begin{equation}
\Sigma'(R){\sigma_p'}^2(R) = \sum_{k=1,2}\int_0^{\infty}
dR'\ \Sigma(R'){\sigma_p}^2(R') \int_0^{2\pi} d\theta\ K_i(R,R',\theta) ,
\label{proj}
\end{equation}
\begin{eqnarray}
K_k(R,R',\theta) = {A_k\pi\sigma_k^2\over 8lw}
& &\left[{\rm erf}\left({R'\cos\theta-R+l\over\sqrt{2}\sigma_k}\right) - 
{\rm erf}\left({R'\cos\theta-R-l\over\sqrt{2}\sigma_k}\right)\right]\times 
\nonumber \\
& &\left[{\rm erf}\left({R'\sin\theta+w\over\sqrt{2}\sigma_k}\right) -
{\rm erf}\left({R'\sin\theta+w\over\sqrt{2}\sigma_k}\right)\right]
\label{conv}
\end{eqnarray}
where $X=R'\cos\theta$ and $Y=R'\sin\theta$.
This relation can be discretized by writing 
$g_i = \Sigma'(R_i){\sigma^2_p}'(R_i)$ and
$h_j = \Sigma(R_j)\sigma^2_p(R_j)$, where $R_i$ and $R_j$ are 
radial grid values.
We then have
\begin{equation}
g_i \approx \sum_{j=1}^{n-1} A_{ij} {h}_j,
\label{matrix1}
\end{equation}
with
\begin{equation}
A_{ij} = \int_{R'_j}^{R'_{j+1}} dR' \int_0^{2\pi} d\theta\ 
K(R_i,R',\theta).
\end{equation}
This is a ``product integration''scheme, and the desired function $h$ 
is obtained by inverting Eq. (\ref{matrix1}); $h_j$ is identified 
with the solution value at the midpoint of the interval $(R_j, R_{j+1}$).

One could attempt to invert Eq. (\ref{matrix1}) directly, but the 
result would be unacceptably noisy due 
to the ill-conditioning of the matrix ${\bf A}$.
(We in fact attempted direct inversion, with discouraging 
results.)
Instead, we looked for the roughness-penalized function $\hat{h}_j$ that 
minimizes
\begin{equation}
\sum_i\left( g_i - \sum_j A_{ij}h_j\right)^2 + 
\lambda\int \left({d^2\log \sigma_p\over dR^2}\right)^2 dR
\label{pen1}
\end{equation}
for some appropriately-chosen $\lambda$, the smoothing parameter.
We represented the penalty function discretely as well, via the
approximation
\begin{equation}
\lambda\sum_j(R_{j+1}-R_j)\left[{{\sigma_p}_{j+1} - 
2{\sigma_p}_j + {\sigma_p}_{j-1}\over 
(R_{j+1}-R_j)^2}\right]^2.
\end{equation}
The selection of the optimal smoothing parameter in 
ill-conditioned problems is, itself, an ill-conditioned problem
(Wahba 1990, p. 105).
An ``infinitely smooth'' estimate, i.e. the limiting estimate for 
large $\lambda$, would be exponential in our case, $\log\sigma_p\propto R$.
We estimated the optimal $\lambda$ by eye.
Appreciably smaller values led to estimates that seemed unphysically 
noisy, while much larger values biased the solution toward 
the infinitely smooth, exponential dependence.
For a fairly wide range of $\lambda$'s around our adopted 
value, however, there was no significant variation in the 
form of the estimated function.

The results are shown in Fig. 1.
The solid curve is our estimate of the corrected, line-of-sight velocity 
dispersion profile, 
$\hat{\sigma}_p(R)=\sqrt{\hat{h}(R)/\Sigma(R)}$.
The dashed lines in Fig. 1 are confidence bands on the estimate.
These were constructed by first generating 300 bootstrap data 
sets $\tilde\sigma_p(R_i)$ from the original data, according to the scheme:
\begin{equation}
\tilde\sigma_p(R_i) = f(R_i) + \delta_i,
\end{equation}
where $f$ is a smoothing spline fit to the uncorrected data and 
$\delta_i \sim {\cal N}(0,\epsilon_i)$ is a random number generated 
from the normal distribution with dispersion equal to the 
measurement error at $R_i$ (Wahba 1990, p. 71).
For each of these bootstrap samples, the deblurring algorithm 
just defined was used to estimate $\sigma_p(R)$.
The distribution of bootstrap estimates at every $R$ was then 
used to construct the confidence bands in the figure.

The corrected velocity dispersion profile in Fig. 1 follows the raw data 
outside of $\sim 1''$.
At smaller radii, the corrected profile rises 
slightly above the mean relation defined by the measured values.
We would not expect the corrected profile to deviate strongly from 
the raw data even at these small radii, since the observed
velocity dispersions are approximately constant within the inner 
two arc seconds, and a flat profile remains flat after deblurring.
However, the bootstrap confidence bands suggest that quite a 
wide range of profile shapes are consistent with the data inside of $\sim 
1''$, from a flat or inwardly decreasing profile, to one that 
increases as far as $\sim 450$ km s$^{-1}$ at $0.1''$ ($95\%$).
We note that the widening of the confidence bands at small radii 
is due primarily to the amplification of errors that accompanies the
deconvolution; the measurement errors quoted by \cite{van94}
are almost independent of radius, but their consequences 
become greater near the center due to the increased importance 
of the seeing and instrumental corrections there.

The flatness of the velocity dispersion profile very near the 
center of M87 was noted by \cite{dre80}, \cite{dre90},
\cite{jar91} and \cite{van94} in their respective
data sets.
This result appears therefore to be robust, although it could 
of course reflect some systematic error in the interpretation
of the data.
\cite{dre90} emphasized the difficulty of reconciling 
a low central velocity dispersion with a massive ($M_{h}> 
10^9M_{\odot}$) nuclear black hole, unless the stellar motions 
become strongly anisotropic near the center.
We will confirm their result below.

\section{Estimation of the Velocity Anisotropy}

In a nonrotating, spherical stellar system, the dependence of the 
velocity anisotropy $\beta=1-\sigma_t^2/\sigma_r^2$ on radius can 
be inferred uniquely from the observed velocity dispersion 
profile $\sigma_p(R)$, if one is willing to assume that the 
functional form of the 
gravitational potential is known (Binney \& Mamon 1982).
We assume that the gravitational potential of M87 can be 
expressed as
\begin{equation}
\Phi(r) = -{GM_{h}\over r} + \left({M\over L}\right) \Phi_L(r),
\label{potent}
\end{equation}
where $M_{h}$ is the mass of the central black hole, $M/L$ is 
the mass-to-light ratio of the stars, and $\Phi_L$ is the 
``potential'' corresponding to the stellar luminosity 
distribution:
\begin{equation}
\Phi_L(r) = -4\pi G\left({1\over r}\int_0^r\nu r^2 dr + 
\int_r^{\infty}\nu r dr\right),
\end{equation}
with $\nu(r)$ the luminosity density.
The expression (\ref{potent}) is not completely general, since it assumes that 
the stellar mass-to-light ratio is independent of radius.
However we expect Eq. (\ref{potent}) to be sufficiently 
general out to a radius of several arc seconds where the black 
hole dominates the gravitational force.
Relaxing the assumption of a constant $M/L$ would make the derivation of 
$\beta(r)$ impossible without additional kinematical information 
(Merritt 1993b).

The stellar velocity dispersion in the radial direction is given 
by
\begin{eqnarray}
\nu(r) \sigma_{r}^{2}(r)=&-&{2 \over \pi r^{3}}
\int_{r}^{\infty} \left[{r \over \sqrt{R^{2}-r^{2}}} +
\cos^{-1} \left({r \over R}\right) \right] \Sigma (R) \sigma_{p}^{2} 
(R)\ RdR \\
& + & {2 \over 3 r^{3}} \int_{r}^{\infty} \left( r'^{3} + {1 \over 2}
r^{3} \right) \nu (r')\ {d \Phi \over dr'}\ dr'
\label{sigr}
\end{eqnarray}
(Dejonghe \& Merritt 1992, Eq. 57a), and the tangential component 
follows from the spherical Jeans equation:
\begin{equation}
\sigma_{t}^{2}=\sigma_{r}^{2} + {r\over 2} \left( {d \Phi \over dr} +
{1 \over \nu} {d\nu \sigma_{r}^{2} \over dr} \right).
\label{Jeans}
\end{equation}
We can obtain a nonparametric estimate of $\sigma_r(r)$ by 
substituting our smooth estimates of $\nu(r)$, $\Sigma(R)$ and 
$\sigma_p(R)$ into Eq. (\ref{sigr}) (Wahba 1990, p. 19).
However we first need to specify the parameters 
$M_{h}$ and $M/L$ that appear in our expression (\ref{potent})
for the potential.

We first consider $M_h$.
\cite{for94} used HST WFPC-2 narrow-band images of M87 to 
find a 100 pc ($\sim 1''$) scale disk of ionized gas with a major 
axis oriented approximately perpendicular to the M87 jet.
\cite{har94} used COSTAR and the FOS to measure the 
velocity at five positions in this disk and concluded that there 
was Keplerian rotation around a mass of $(2.4\pm 0.7)\times 10^9 
M_{\odot}$.
More recent HST observations with a smaller, $0.086''$ aperture 
appear to confirm this interpretation, although the 
scatter in the velocities is still consistent with a fairly wide 
range of masses, from $1$ to $3.5\times 10^9 M_{\odot}$ (Ford et al. 
1996).
This range is essentially equal to the $\pm 2\sigma$ interval 
quoted by \cite{har94}.
We will accordingly compute our kinematical solutions using 
the three values $M_{h}=\{1.0, 2.4, 3.8\}\times 10^9 M_{\odot}$, 
corresponding to the \cite{har94} value 
plus or minus twice its estimated uncertainty.

The second parameter, $M/L$, can be derived from the observed 
velocity dispersion profile independent of
any assumptions about the stellar velocity anisotropy.
The virial theorem
\begin{equation}
3\int_0^{\infty}\Sigma\sigma_p^2R\ dR = 
2\int_0^{\infty}\nu{d\Phi\over dr}r^3\ dr
\end{equation}
(e. g. Dejonghe \& Merritt 1992, Eq. 51)
becomes, with our assumed form for $\Phi(r)$,
\begin{equation}
{M \over L}= {3 \int_0^{\infty} dR\ R \Sigma \sigma_p^2
-2G M_{h} \int_0^{\infty} dr\ r \nu \over 8\pi G\int_0^{\infty} 
dr\ r\nu \int_0^r dr'r'^2\nu},
\label{moverl}
\end{equation}
which allows $M/L$ to be determined from previously-computed 
quantities.
(In order to evaluate the integrals in Eqs. (\ref{sigr}) and (\ref{moverl}) 
that extend to infinite radius, the velocity dipsersion profile 
was extrapolated as a power law beyond $25''$.)
Our estimates of the I-band $M/L$ in solar units are given in Table 1.
Finally, confidence bands on $M/L$ and on the function estimates 
were generated by repeating the set of computations just 
described for each of the $300$, bootstrap-generated data sets 
defined above.

Estimates of $\sigma_r(r)$, $\sigma_t(r)$ 
and $\beta(r)$ and their 95\% confidence bands are shown in Fig. 
2.
Fig. 2b gives results for $M_{h}=2.4\times 10^9 M_{\odot}$, the 
value preferred by \cite{har94},
while Figs. 2a and 2c illustrate how these function estimates 
change when $M_{h}$ is increased or decreased by
twice its estimated uncertainty of $0.7\times 10^9 M_{\odot}$.
The confidence bands become very wide at both large and small
radii.
At small radii, the effects of spatial deprojection and instrumental
deblurring are greatest.
At large radii, the lack of velocity data beyond $\sim 25''$ implies
a large variance in quantities like $\sigma_r$ that formally depend
on integrations to infinity.

For $M_{h}=1.0\times 10^9M_{\odot}$, $\beta(r)$ is found to 
remain approximately constant with radius, $\beta(r) \approx 
0.4-0.5$, corresponding to $\sigma_r/\sigma_t\approx 1.3\sigma_t$.
The stellar velocity ellipsoid is radially 
elongated everywhere in this case, although the 95\% confidence 
bands are consistent with motions that are close to isotropic.
The need for radially-biased motions when the black hole mass is 
small or zero has been noted by a number of authors including 
\cite{bin82}, \cite{dre90}, and \cite{van94}.

However when $M_{h}$ is increased to its preferred value of 
$2.4\times 10^9 M_{\odot}$, $\beta(r)$ is forced to decline at small 
radii in order to maintain the observed, low value of the central 
velocity dispersion.
The central decline in $\beta(r)$ becomes even more extreme if 
$M_{h}$ is increased to $3.8\times 10^9 M_{\odot}$ -- in this 
case, the inferred value of $\sigma_r^2$ actually falls below 
zero inside of $\sim 0.4''$ (Fig. 2c).
A similar conclusion was reached by \cite{dre90},
who used an orbit-based code to infer the stellar 
kinematics in M87 assuming $M_{h}= 3.6\times 10^9 M_{\odot}$.

Although our best estimates of $\beta(r)$ suggest a rapidly 
increasing anisotropy at small radii, the confidence bands on 
$\beta$ become very wide at small $r$, and it is not 
obvious from Fig. 2 whether the central 
decline in $\beta$ that we find using the two larger values of 
$M_{h}$ is statistically significant.
One way to address this question is to ask: What fraction of the 
bootstrap estimates have a lower value of $\beta$ at $0.5''$ 
(say) than at $2''$?
We find that the decrease in $\beta$ between $2''$ and $0.5''$ 
is significant at the $98\%$ level for $M_{h}=2.4$ and 
$3.8\times 10^9 M_{\odot}$, and at the $43\%$ level for 
$M_{h}=1.0\times 10^9 M_{\odot}$.
If \cite{for96} are correct in speculating that turbulent 
motions in the M87 gas disk might imply an even larger mass than 
$3.8\times 10^9 M_{\odot}$, the central decline in $\beta(r)$ would 
become even more significant.

\section{Discussion}
Our conclusion that $\sigma_r$ is approximately equal to, or 
slightly greater than, $\sigma_t$ throughout the main body of M87 
is unremarkable.
However the apparent predominance of circular over radial motions 
at the very center of M87 is potentially more interesting.
A number of models for the formation of nuclear black holes 
predict a bias toward circular motions within the radius of 
influence of the black hole, $r_{h} = GM_{h}/\sigma^2$, or $\sim 1''$ 
in M87.
This is roughly the radius at which we find that $\sigma_r$ must
fall below $\sigma_t$ when $M_{h} \ge 1.0\times 10^9 M_{\odot}$.
(As often noted, $r_{h}$ in M87 is also roughly equal to the size of 
the seeing disk, a coincidence that is hopefully fortuitous!)
Thus it seems appropriate to ask whether the anisotropy that we 
infer in the stellar motions could be a relic of the black hole 
formation process.

In one widely-discussed class of models, black holes grow from the 
accumulation of gas on a timescale long enough that the stellar 
action variables are adiabatically conserved.
The gradual deepening of the gravitational potential in these 
models can induce a tangential anisotropy in the stellar 
motions at radii near $r_{h}$.
\cite{you80} first noticed this effect in numerical calculations, 
and \cite{goo84} showed that the adiabatic growth of a black 
hole at the center of an initially isothermal galaxy always 
enhances the tangential components of the velocity dispersion 
at the expense of the radial component.
However the effect is subtle, since the eccentricities of 
orbits in axisymmetric systems are almost unaffected by slow 
changes in the potential (Lynden-Bell 1963).
\cite{qui95} followed the adiabatic growth of black holes in a 
variety of spherical, initially isotropic galaxy models, with
and without constant-density cores.
The resulting $\beta(r)$ profiles varied from model to 
model, but the value of $\beta$ at $r=r_{h}$ was always close to
its minimum, with $\beta_{min}$ typically lying between $-0.3$ and $0$.
Such mild anisotropies would be almost undetectable 
in M87 given the quality of data currently available.

Black holes might also grow by the capture of stars on low angular 
momentum orbits, whose pericenters lie within the tidal breakup 
radius $r_t$ of the black hole (Hills 1975).
One result would be to remove elongated orbits from the central 
regions of the galaxy and hence to increase the ratio of $\sigma_t$ 
to $\sigma_r$ (e.g. Cohn \& Kulsrud 1978, Fig. 8).
However the tidal radius of a $\sim 10^9 M_{\odot}$ black hole is 
comparable to its Schwarzschild radius, or $\sim 10^{-4}$ pc in the case of 
the M87 black hole -- much too small to be observed.
Scattering of stars at larger radii into the loss cone (Frank \& Rees 1976) 
would be unimportant due to the long relaxation time in M87.

In yet another class of models, nuclear black holes grow through the
accretion of other black holes acquired from galaxy mergers 
(e.g. Ebisuzaki, Makino \& Okumura 1991). 
Dynamical friction drags the black holes into the central region 
where they form a bound pair.
The subsequent evolution was outlined by \cite{beg90}:
the binary separation shrinks with time---at first through dynamical 
friction, later because of three-body scattering processes in 
which the hard binary ejects stars from the center---until it becomes 
small enough that gravitational radiation causes the black holes to merge.  
The mass of stars ejected by the binary can be comparable to the sum of 
the two black hole masses (Quinlan 1996).
Because the binary interacts most strongly with stars on elongated 
orbits, three-body scattering should introduce a tangential bias in the 
velocity distribution of the remaining stars.
Detailed simulations of this complicated process have yet
to be published, but it seems plausible that the induced anisotropy could
be much greater than in the adiabatic-growth or tidal-disruption models.

Although our conclusions about the stellar dynamics of M87 are
qualitatively consistent with the predictions of all these models,
we stress that the case for a massive black hole is not 
strengthened in any way by our work.
The low, central value of the stellar velocity dispersion in M87
continues to be surprising given the independent evidence for a 
dark mass.
If a massive black hole is indeed present in M87, then nature 
has somehow conspired to hide its effects on the stellar kinematics 
by adjusting the shapes of the stellar orbits.
As discussed above, such a conspiracy is not impossible to imagine, 
but one would still like to find evidence in the stellar 
kinematics for the central singularity.
The bias toward circular orbits that we infer at $\sim 1''$
is not sufficient to keep the observed dispersions low at
arbitrarily small radii; eventually, even exactly circular motions
would be expected to show a Keplerian rise.
Observations of M87 with the Space Telescope Imaging Spectrograph 
(Baum et al. 1996) should allow accurate measurement of the stellar 
velocity dispersion at a radius of $0.2''\approx 16$ pc, where the 
orbital velocity around a $2.4\times 10^9 M_{\odot}$ black 
hole would be $\sim 800$ km s$^{-1}$.
Such high velocities would leave their mark on the 
line-of-sight velocity dispersion profile regardless of the 
shapes of the stellar orbits.

\bigskip\bigskip
We thank R. van der Marel for assisting in the data 
interpretation, and G. Quinlan for helping us to understand the
various models for black hole formation.
This work was supported by NSF grant AST 90-16515 and NASA grant 
NAG 5-2803 to DM.

\begin{table}
\caption{M87 Stellar Mass-To-Light Ratio}
\begin{tabular}{cccc}
\\ \hline
$M_{h}/M_{\odot}$ & $(M/L)_I$ & $68\%$ & $95\%$ \\ \hline
$1.0\times 10^9$ & $3.76$ & $(2.94, 5.96)$ & $(2.33, 9.44)$  \\
$2.4\times 10^9$ & $3.72$ & $(2.89, 5.91)$ & $(2.29, 9.39)$  \\
$3.8\times 10^9$ & $3.68$ & $(2.86, 5.88)$ & $(2.26, 9.31)$  \\ \\
\end{tabular}
\end{table}

\clearpage

\figcaption[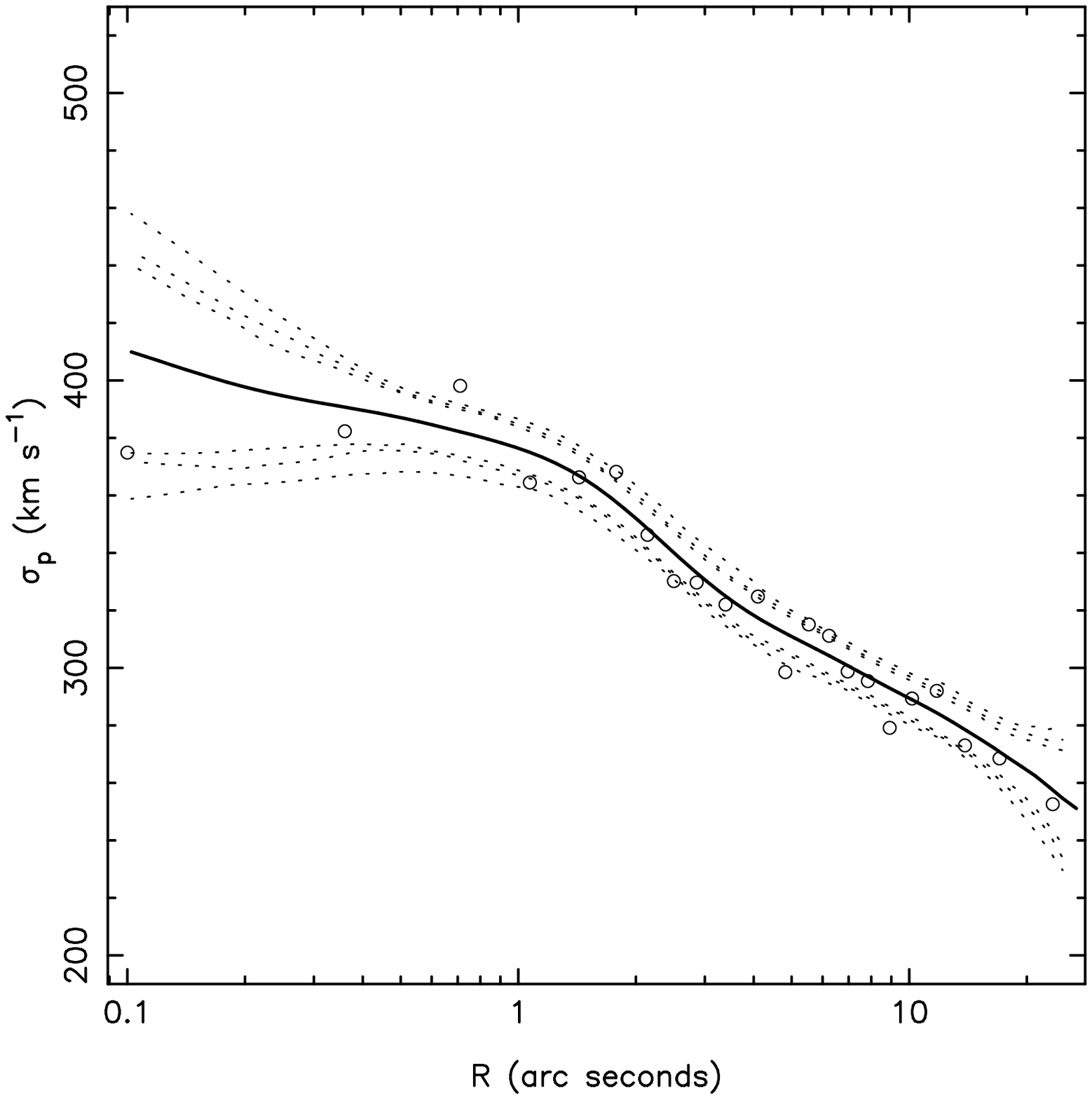]{\label{fig1}} Line-of-sight velocity 
dispersions in M87.
Open circles: G-band measurements from \cite{van94}.
The innermost data point has been shifted from its true position
at $R=0''$.
Solid line: velocity dispersion profile corrected for the effects of
seeing and instrumental blurring.
Dashed lines: 90\%, 95\% and 99\% confidence bands on the corrected
profile.

\figcaption[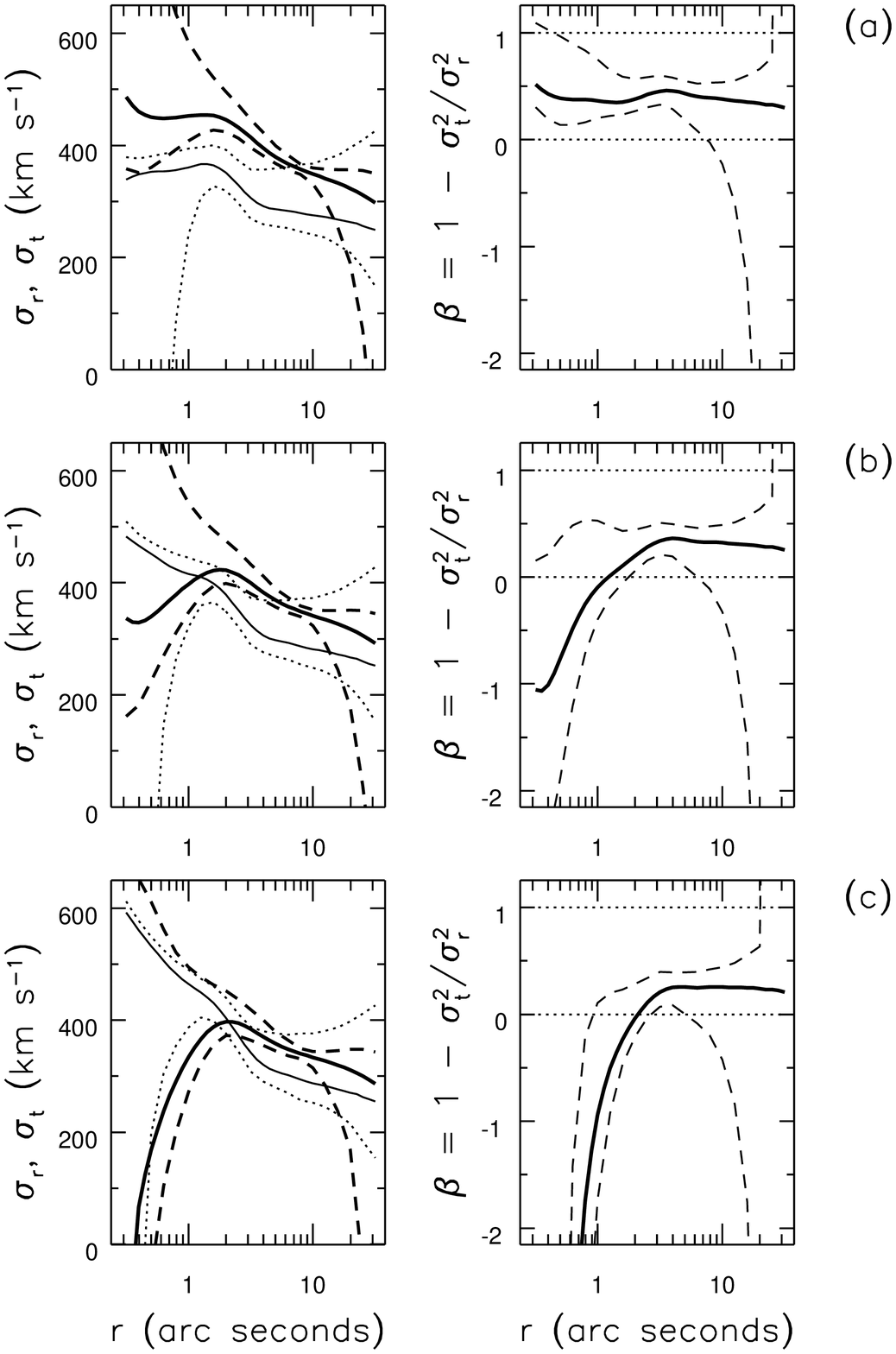]{\label{fig2}} Velocity dispersions and 
anisotropy as a function of radius in M87, for three different
values of the black hole mass: $M_{h}=1.0\times 10^9 M_{\odot}$ (a),
$2.4 \times 10^9 M_{\odot}$ (b) and $3.8\times 10^9 M_{\odot}$ (c).
Left hand column: radial (heavy curves) and transverse (thin curves)
velocity dispersions; dashed curves are 95\% confidence bands.
Right hand column: velocity anisotropy, 
$\beta=1-\sigma_t^2/\sigma_r^2$, with 95\% confidence bands.

\setcounter{figure}{0}

\begin{figure}
\plotone{figure1.ps}
\caption{ }
\end{figure}

\begin{figure}
\plotone{figure2.ps}
\caption{ }
\end{figure}


\clearpage

\begin{thebibliography}{}

\bibitem[]{}

\bibitem[Baum, S. et al. 1996]{bau96} Baum, S. et al. 1996, STIS
	Instrument Handbook, Version 1.0 (Baltimore: STScI)
\bibitem[Begelman, Blandford \& Rees (1990)]{beg90} Begelman, M. C., 
	Blandford, R. D. \& Rees, M. J. 1980, Nature, 287, 307
\bibitem[Binney \& Mamon (1982)]{bin82} Binney, J. J. \& Mamon, G. A. 1982,
	MNRAS, 200, 361
\bibitem[Cohn \& Kulsrud 1978]{coh78} Cohn, H. \& Kulsrud, R. M. 1978,
	ApJ, 226, 1087
\bibitem[Dejonghe \& Merritt 1992]{dej92} Dejonghe, H. \& 
	Merritt, D. 1992, ApJ, 391, 531
\bibitem[Dressler (1980)]{dre80} Dressler, A. 1980, ApJ, 240, L11
\bibitem[Dressler \& Richstone (1990)]{dre90} Dressler, A. \& 
	Richstone, D. O. 1990, ApJ, 348, 120
\bibitem[Ebisuzaki, Makino \& Okumura (1991)]{ebi91} Ebisuzaki, T., 
	Makino J., \& Okumura, S. K. 1991, Nature, 354 212
\bibitem[Efron (1982)]{efr82} Efron, B. 1982, The Jackknife, the
	Bootstrap and Other Resampling Plans (Philadelphia: SIAM)
\bibitem[Ford et al. (1994)]{for94} Ford, H. C., Harms, R. J., 
	Tsvetanov, Z. I., Hartig, G. F., Dressel, L. L., Kriss, G. A., 
	Bohlin, R. C., Davidsen, A. F., Margon, B. \& Kochhar, A. K. 
	1994, ApJ, 435, L27
\bibitem[Ford et al. (1996)]{for96} Ford, H. C., Ferrarese, L., Hartig, G.,
	Jaffe, W., Tsvetanov, Z. \& van den Bosch, F. 1996, in 
	Nobel Symposium No. 98, Barred Galaxies and Circumnuclear
	Activity, ed. A. Sandqvist \& P. O. Lindblad (Heidelberg:
	Springer), 293
\bibitem[Frank \& Ress 1976]{fre76} Frank, J. \& Rees, M. J.1976, 
	MNRAS, 176, 633
\bibitem[Goodman \& Binney (1984)]{goo84} Goodman, J. \& Binney, J. J.
	1984, MNRAS, 207, 511
\bibitem[Harms et al. (1994)]{har94} Harms, R. J., Ford, H. C., 
	Tsvetanov, Z. I., Hartig, G. F., Dressel, L. L., Kriss, G. A., 
	Bohlin, R. C., Davidsen, A. F., Margon, B. \& Kochhar, A. K. 
	1994, ApJ, 435, L35
\bibitem[Hills (1975)]{hil75} Hills, J. G. 1975, Nature, 254, 295
\bibitem[Jarvis \& Peletier (1991)]{jar91} Jarvis, B. J. \& 
	Peletier, R. F. 1991, AA, 247, 315
\bibitem[Kormendy \& Richstone 1995]{kor95} Kormendy, J. \& 
	Richstone, D. O. 1995, ARAA, 33
\bibitem[Lauer et al. (1992)]{lau92} Lauer, T. R. et al. 1992, 
	AJ, 103, 703
\bibitem[Lynden-Bell 1963]{lyn63} Lynden-Bell, D. 1963, The Observatory,
	No. 932, 23	
\bibitem[Merritt (1987)]{mer87} Merritt, D. 1987, ApJ, 319, 55
\bibitem[Merritt (1993a)]{mer93a} Merritt, D. 1993a, in Structure,
	Dynamics and Chemical Evolution of Elliptical Galaxies,
	ed. I. J. Danziger, W. W. Zeilinger \& K. Kj\"ar (Munich: ESO),
	275 
\bibitem[Merritt (1993b)]{mer93b} Merritt, D. 1993b, ApJ, 413, 79
\bibitem[O'Sullivan (1986)]{osu86} O'Sullivan, F. 1986, Statist. 
	Sci., 1, 502 
\bibitem[Peletier et al. (1990)]{pel90} Peletier, R. F., Davies, R. 
	L., Illingworth, G. D., Davies, L. E. \& Cawson, M.
	1990, AJ, 100, 1091
\bibitem[Qian et al. (1995)]{qia95} Qian, E. E., de Zeeuw, P. T.,
	van der Marel, R. P. \& Hunter, C. 1995, MNRAS, 274, 602 
\bibitem[Quinlan (1996)]{qui96} Quinlan, G. D. 1996, New Astronomy, 1, 
	35
\bibitem[Quinlan, Hernquist \& Sigurdsson (1995)]{qui95} Quinlan, 
	G. D., Hernquist, L. \& Sigurdsson, S. 1995, ApJ, 440, 554
\bibitem[Richstone \& Tremaine (1985)]{ric85} Richstone, D. O. \&
	Tremaine, S. 1985, ApJ, 296, 370
\bibitem[Sargent et al. (1978)]{sar78} Sargent, W. L. W., Young, P. 
	J., Boksenberg, A., Shortridge, K., Lynds, C. R. \&
	Hartwick, F. D. A. 1978, ApJ, 221, 731.
\bibitem[van der Marel (1994)]{van94} van der Marel, R. P. 1994, MNRAS,
	270, 271
\bibitem[Wahba (1990)]{wah90} Wahba, G. 1990, Spline Models for
	Observational Data (Philadelphia: SIAM)
\bibitem[Young (1980)]{you80} Young, P. 1980, ApJ, 242, 1232

\end{thebibliography}
\end{document}